\documentstyle[preprint,aps,epsfig]{revtex}
\tightenlines
\begin{document}
\draft
\title{
Critical behavior of frustrated systems:
Monte Carlo simulations versus Renormalization Group}
\author{D. Loison$^{1}$, A. I. Sokolov$^{2,3}$, B. Delamotte$^4$,
S. A. Antonenko$^2$, K. D. Schotte$^1$ and H. T. Diep$^5$}
\address{
$^1$ Institut f\"ur Theoretische Physik, Freie Universit\"at Berlin,
Arnimallee 14, 14195 Berlin, Germany\\
$^2$ Department of Physical Electronics, Saint Petersburg Electrotechnical
University, Professor Popov Street 5, St. Petersburg 197376, Russia\\
$^3$ Department of Physics, Saint Petersburg Electrotechnical
University, Professor Popov Street 5, St. Petersburg 197376, Russia\\
$^4$ Laboratoire de Physique Th\'eorique et Hautes Energies,
Universit\'e Paris 7, 2 place Jussieu, 75251 Paris Cedex 05, France\\
$^5$ Laboratoire de Physique Th\'eorique et Mod\'elisation,
Universit\'{e} de Cergy-Pontoise,
2, Avenue Adolphe Chauvin, 95302 Cergy-Pontoise Cedex, France\\
Damien.Loison@physik.fu-berlin.de, ais@sokol.usr.etu.spb.ru,
delamotte@lpthe.jussieu.fr, schotte@physik.fu-berlin.de, diep@u-cergy.fr
}
\maketitle
\begin{abstract}
We study the critical behavior of frustrated systems   by means of
Pade-Borel resummed three-loop renormalization-group expansions
and numerical Monte Carlo simulations.
Amazingly, for six-component spins where the transition is second order,
both approaches disagree.
This unusual situation is analyzed both from the point of view of the
convergence of the resummed series and from the possible relevance of non
perturbative effects.
\end{abstract}
\vspace{1.cm}
P.A.C.S. numbers:05.50+q,05.70.Fh, 75.10.Hk,64.60.Cn, 75.10.-b

Frustrated spin systems have been very much studied in their classical as
well as quantum aspects. In particular, the critical behavior of 3$D$
Stacked Triangular Antiferromagnets (STA) has deserved much attention
\cite{Diep2,Delamotte,ASV,AS94,CP97,Zumbach93,Kawa,LoisonSchotteXY,LoisonSchotteHei}
since, firstly it has many physical realizations in rare earth materials,
secondly it is an archetype for frustrated systems, and thirdly it is
directly related to the behavior of its two dimensional zero temperature,
quantum counterpart. The frustration in this system comes from the fact
that -- for $N > 1$ component spin systems -- the ground state is non
collinear and shows the famous 120$^0$ structure. It is thus natural to
believe that if the transition is second order, it belongs to a new
universality class. Our present understanding of these systems comes as
usual from theoretical renormalization group (RG) calculations, from
Monte Carlo simulations and from experiments. The most impressive fact
concerning these systems is that more than twenty years after the first
works devoted to their study, there is still no agreement between these
approaches. For instance, a calculation made in $D = 4 - \epsilon$
\cite{ASV} predicts no stable fixed point for $N$ in the interval:
$N_{c2} = 2.202 - 0.569 \epsilon + 0.989 {\epsilon}^2 < N < N_{c3}
= 21.80 - 23.43 \epsilon + 7.088 {\epsilon}^2$
and another made in $D = 2 + \epsilon$ predicts a fixed
point for any $N$ larger than 2. Some experiments find a second order phase
transition while others a weak first order. Moreover, the different approaches
finding a second order transition do not find the same exponents, a fact
that suggests that the theoretical or numerical approaches may miss a
fundamental point (presence of topological defects, breakdown of
perturbation theory, lattices of too small sizes\dots).

Our aim in this Letter is to shed light on this problem. We rely on the
fact that the three-loop RG calculations made in $D = 4 - \epsilon$  with
$\epsilon=1$ and
directly in $D = 3$ find a critical value $N_c(D=3)$
above which the transition is second order  equal to 3.39 \cite{ASV} and
3.91 \cite{AS94}, respectively.
Below it, it is supposed to be first order.
We therefore expect a very weakly first order transition for $N=3$ (and even
perhaps for $N=4$) -- a situation very difficult to test numerically.
Therefore, instead of studying directly the physical, i.e. $N = 3$, spin
system, we choose to study the following question: is there consensus
between the results given by the usual RG approach based on the
Landau-Wilson ($\phi^4$ - like) model and those obtained by Monte Carlo
simulations for the values of $D$ and $N$ where a fixed point is found?

Let us emphasize that the reliability of the RG approach for
predicting the three-dimensional critical behaviors is not generic
but has been demonstrated for most popular and simplest universality
classes such as the usual $O(N)$ and cubic ones. The
discrepancy between the perturbative
results around $d=2$ and $d=4$ is in fact common to a wide class of
systems among which the dipole locked phase of $^3$He, electroweak phase
transition, smectic liquid crystal, etc. Our study is therefore likely to
be relevant to a much wider class of systems than the frustrated magnets.

To tackle with our question, we study in $d = 3$ the largest possible $N$
compatible with numerical possibilities where the usual recipes should
work since in this case we are far above the line $N_c(D)$, the proximity
of which could be the root of all the problems. Being in principle in
the second order region, we expect to compute accurately the critical
exponents both numerically and in a RG approach using the usual resummation
techniques. The comparison between the results obtained by these two
methods should be a test of the most powerful theoretical approaches
in this non-ferromagnetic case. We also choose the value of $N$ such that the
corresponding system does not show topological defects in order to eliminate
a possible reason for the breakdown of perturbation theories. It turns out
that $N = 6$ is the ideal candidate. We present in this article numerical
results for $N=6$ as well as analytical ones for many $N$ and compare them.

{\sl Renormalization Group calculations}.
Let us first show our results obtained from the renormalization group
analysis. The Landau-Wilson Hamiltonian relevant to the STA system is:
\begin{equation}
\label{hamiltonien}
H={1 \over 2} \int d^3x \Bigg{[} r_0^2 \phi_\alpha \phi_\alpha^*+
\nabla \phi_\alpha \nabla\phi_\alpha^* +
{u_0 \over 2} \phi_\alpha \phi_\alpha^* \phi_\beta \phi_\beta^* +
{w_0 \over 2} \phi_\alpha \phi_\alpha \phi_\beta^* \phi_\beta^* \Bigg{]}.
\end{equation}
The domain of parameters of interest is $u_0 > 0$ and $w_0 > 0$. The
calculations are based on the three-loop RG equations obtained earlier
for the more complicated model having three independent quartic coupling
constants \cite{AS94}. They are carried out directly in $d = 3$
and a Pad\'e-Borel resummation of the relevant expansions is performed.
Pad\'e approximants [3/1] and [2/1] are employed for analytical continuation
of the Borel transform series for the $\beta$-functions and critical
exponent $\gamma$, respectively. The Fisher exponent is evaluated by direct
substitution of the fixed point coordinates into the corresponding expansion.

The fixed point of the RG flow diagram which controls the non-trivial (chiral)
critical behavior is found. For $N > 7$ it turns out to be a stable node,
for $N = 5, 6, 7$ this point manifests itself as a stable focus.
The latter scenario looks quite new, i. e. is observed for the first
time in STA systems, while the former one has been already discussed
(see, e. g., \cite{AS94}). The estimates of critical exponents
$\gamma$ and $\eta$ for various $N$ are obtained from corresponding RG
series. Making use of well-known scaling relations yields numerical
values of the others. The results of our RG calculations are collected
in Table 1 and presented, along with the other data, in Fig. 3.
As is seen, critical exponents as functions of $N$ demonstrate a cusp
between $N = 7$ and $N = 8$ that reflects the abovementioned change
of type of the fixed point governing the critical behavior.

\begin{figure}

\vskip 1cm
\centerline{
\psfig{figure=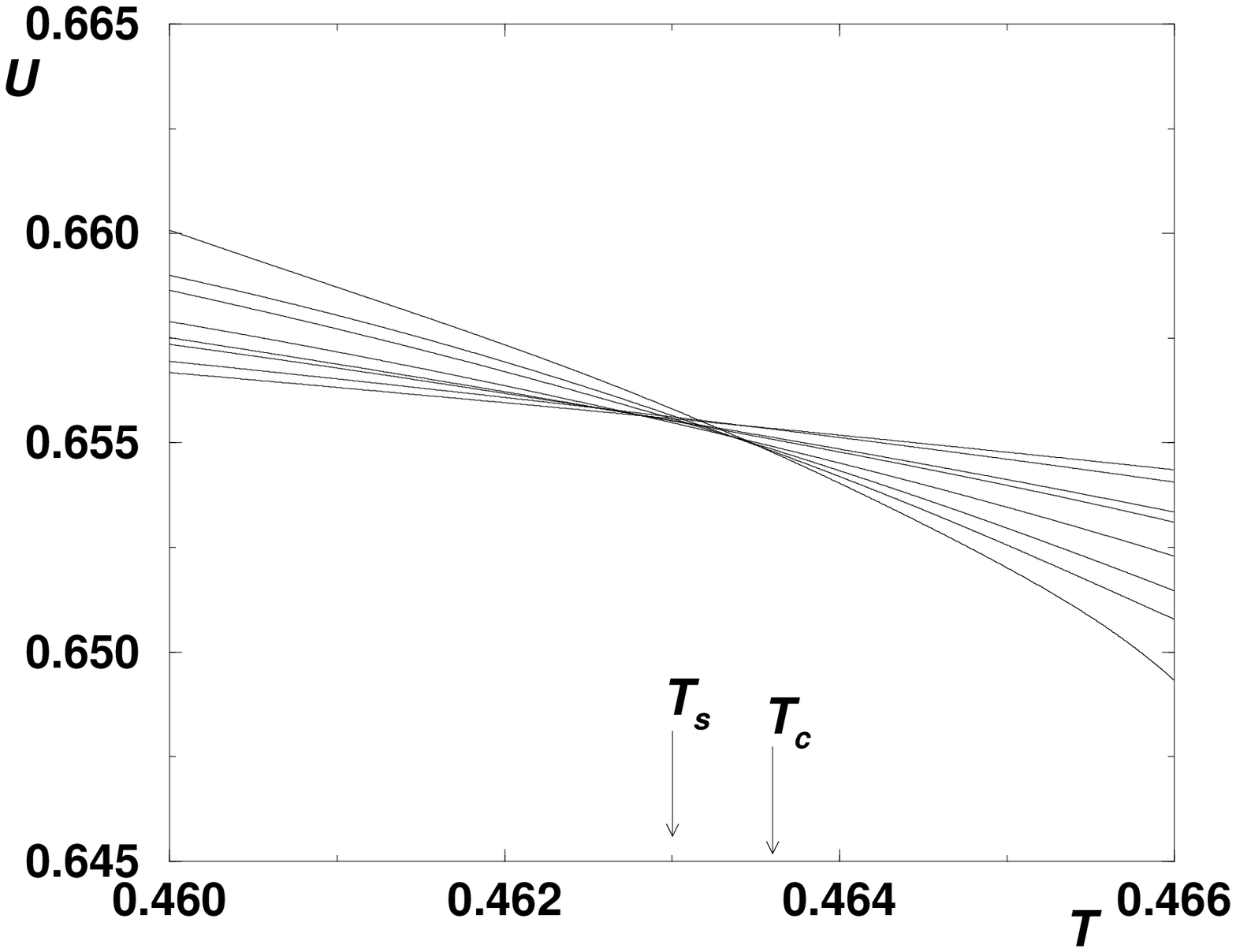,width=6.5cm} }
\vskip 1cm
\caption{\label{fig1}
Binder's parameter $U$
as function of the temperature for different
sizes $L$ (in the left part of the figure from down to up
$L$=12,\,15,\,18,\,21,\,24,\,27,\,30,\,36).
The arrows show the estimated critical temperature $T_c$ and the temperature
of our simulations $T_s$.
}

\vskip 1cm
\centerline{
\psfig{figure=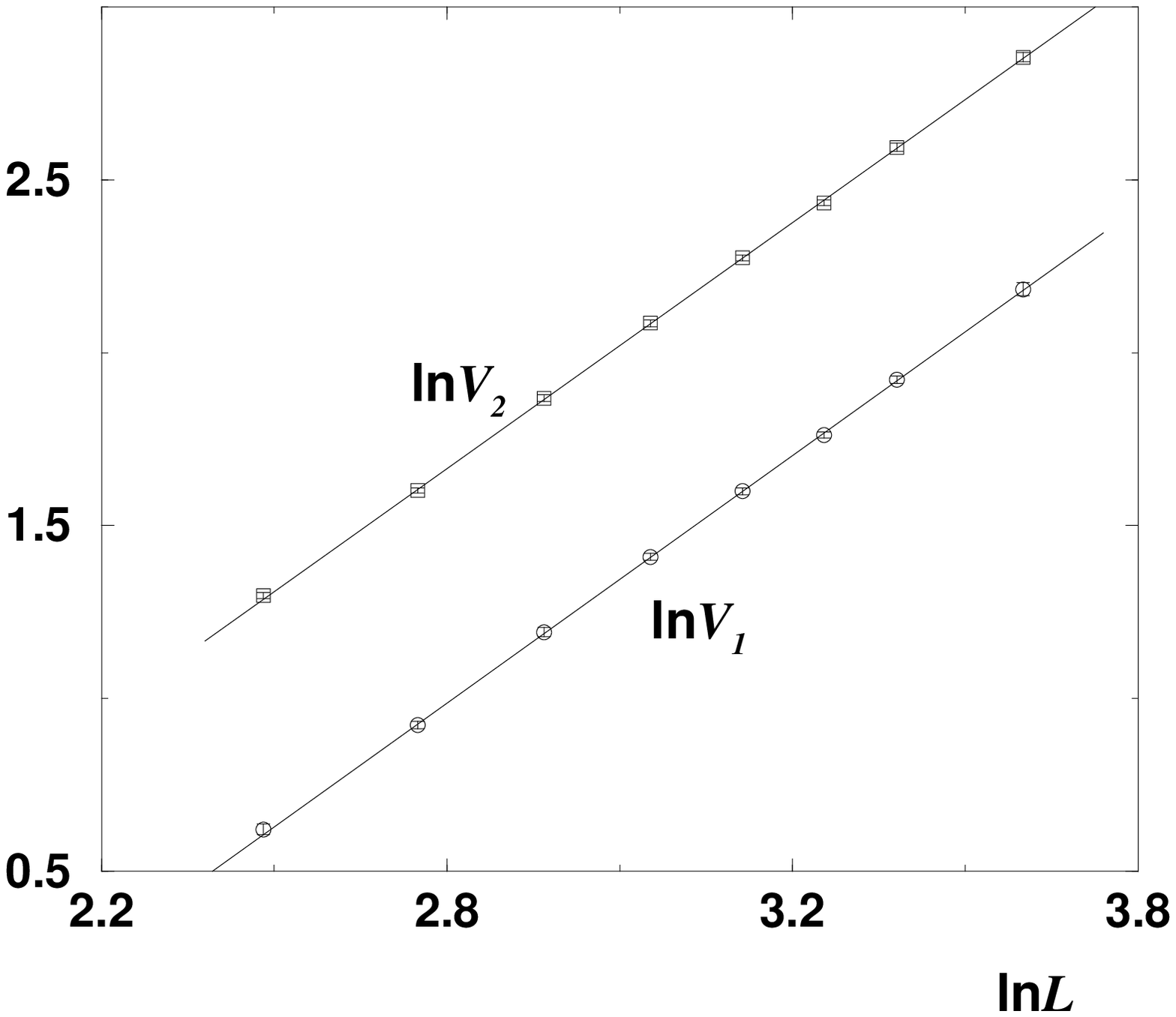,width=6.5cm} }
\vskip 1cm
\caption{\label{fig2}
Values of $V_1$ and $V_2$ as function of $L$ in a ln-ln scale at $T_c$.
The value of the slopes gives $1/\nu$ and we obtain $\nu$=
0.698(12) for  $V_1$ and 0.702(13) for $V_2$. The smallest size ($L=12$)
is not included in our fits.
}

\vspace{0cm}

\vskip 1cm
\centerline{
\psfig{figure=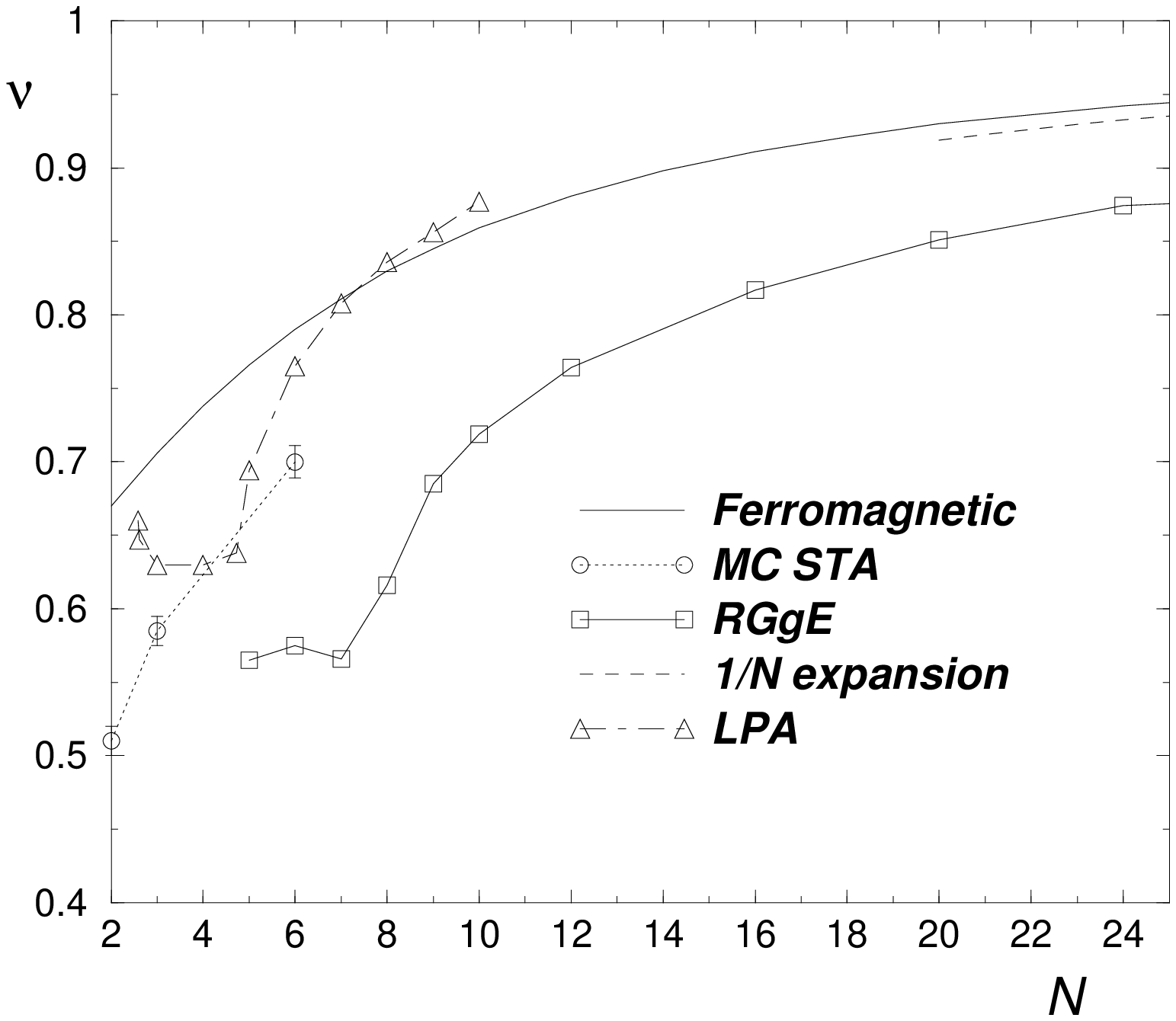,width=6.5cm} }
\vskip 1cm
\caption{\label{fig3}
$\nu$ for different methods (see text).
}
\end{figure}

{\sl Monte Carlo results}.
We present now our MC simulations. We use six-component spins interacting
via the Hamiltonian
\begin{equation}
\label{eq1}
        H = \sum_{(ij)} J_{ij} {\bf S}_{i}.{\bf S}_{j} \ ,
\end{equation}
where the sum runs over all neighbors of the stacked triangular
lattice (STA) and the interaction is chosen antiferromagnetic ($J>0$).
In the ground state the spins are planar with the three
spins at the corners of each triangle forming a 120$^{\circ}$ structure.
We use the standard Metropolis algorithm in combination with the
over-relaxation algorithm \cite{Overrelaxation}. Between each Metropolis
we use one over-relaxation step. This allows us to reduce the correlation
time and obtain better statistics. For each size we use some hundred
thousand steps to equilibrate our system and up to five millions steps
to thermalize for the bigger sizes. We have repeated these simulations
for different initial configurations (ordered or random) to make sure
that our results do not depend on them. We use the histogram MC technique
developed by Ferrenberg and Swendsen \cite{Ferren88}.
 From a simulation done at $T_{0}$, this technique allows us to obtain
thermodynamic quantities at $T$ close to $T_{0}$.
We have studied our system in the finite size scaling (FSS)
region \cite {Barber 83}
and our simulations have been done at $T_s=0.463$.
We consider $L^2*(2 L/3)$ systems, where $(L)^2$ is the size of the planes,
and $2 L/3$ is the number of planes in the $z$ axis. First, to find the
critical temperature $T_c$, we use Binder's cumulant defined as
\begin{eqnarray}
U=1-<M^{4}>/(3<M^{2}>^{2})
\end{eqnarray}
where the order parameter
$M$ is calculated in partitioning our lattice in three sublattices
with only collinear spins and by summing each magnetization.
We record the variation of $U$ with $T$ for various system sizes and then
locate the intersection of these curves.
We compare the values of $U$ for two different lattice sizes $L$
and $L'=bL$, making use of the condition \cite {BinderU}
\begin{eqnarray}
{\frac{U_{bL}}{U_{L}}}\Bigg\arrowvert _{T=T_{c}} = 1.
\end{eqnarray}
In Fig.~\ref{fig1}, $U$ is plotted as a function of the temperature
for different sizes from $L = 12$ up to $L = 36$.
Due to the presence of residual corrections to finite size scaling,
one actually needs to extrapolate the results of this method for
(ln $b$)$^{-1} \rightarrow  0$. From these data,
we extrapolate the value of $T_{c}$ (not shown) and obtain:
$T_{c}=0.4636(2)$ .
We estimate the universal quantity $U$ at $T_{c}$ ($U^*$) as
$U^*=0.6545(15)$.
With the value of $T_c$ we calculate the critical exponents
using log-log fit \cite{Barber 83,Ferren91}.
We obtain from $V_1= <M E>/<M>-<E>$,
$V_2=<M^2E>/<M^2>-<E>$,
(Fig.~\ref{fig2}),  estimates of $1/\nu$,
from the suceptibility $\chi=N<M^{2}>/(k_{B}T)$ (not shown)
estimates of $\gamma/\nu$,
and from $<M>$ (not shown) an estimate of $\beta/\nu$.
Combining these results we obtain respectively,
$\nu=0.700(11)$,
$\gamma/\nu=1.975(20)$, and
$\beta/\nu=0.513(12)$.
All our errors are calculated with the help of the Jackknife procedure
\cite{Jackknife} and include the influence of the uncertainty
in estimating $T_c$.
The results are summarized in Table \ref{table2} where the value of $\eta$ has
been calculated with the hyperscaling relation $\gamma/\nu=2-\eta$. We note
that, contrary to spins with two or three components, $\eta$ is positive.
This is due to the fact that for $N = 6$ the Renormalization Group flow
is attracted by a true stable fixed point and not simply by a local
minimum \cite{Zumbach93,LoisonSchotteXY,LoisonSchotteHei}.

{\sl Discussion}.
The predictions of the renormalization-group $g$-expansion technique
for $N = 6$-component spins listed in Table \ref{table1} do not agree
with the Monte Carlo results given in Table \ref{table2}.
We have plotted in figure \ref{fig3} the results for $\nu$
from the MC data, the RG $g$-expansion (this work), the Local Potential
Approximation method (LPA) \cite{Zumbach93} and the $1/N$ expansion
\cite{Kawa88} (first order).
The six-loop RG results for the ferromagnetic case \cite{AS95} are also
plotted for comparison. The interesting point is that these results
are obtained by methods representing the state of the art
in the field of critical phenomena. When applied to the systems
belonging to the $O(N)$ universality classes they indeed fit very
well together. Let us emphasize that our numerical results
are well converged and it seems unlikely that a non trivial
behavior shows up at much larger system sizes THAT would resolve the
discrepancy with the RG $g$-expansion results.

In order to clear up what can be an origin of the marked discrepancy,
we analyze the structure of the RG expansions employed.
The main attention is paid to the vicinity of
the chiral fixed point of the RG flow for $N = 5, 6, 7$ when
this point is a focus. Contrary to the (unstable) fixed point governing
the $O(N)$-symmetric critical behavior, the chiral point lies very close to
the $w$ axis being far from the  $u$ axis. In particular,
for the case of interest $N = 6$ its coordinates are: $u^* = 0.0665,
w^* = 1.6025$. In this region, the structure of the series of the
$\beta$-functions turns out to be unexpectedly irregular. As an example,
we present here two ``cuts" of the Borel-transformed expansion for
$\beta_u (u, w)$ running through the chiral fixed point which clearly
demonstrate such irregularity:
\begin{eqnarray}
\label{rgexp}
\beta_u^B (u, 1.6025) &=& -0.3607 + 0.7774 u - 0.5004 u^2 + 0.0339 u^3
- 0.0055 u^4,  \\
\label{rgexp2}
\beta_u^B (0.0665, w) &=& 0.0643 - 0.0132 w - 0.1960 w^2 + 0.0346 w^3
+ 0.0010 w^4.
\end{eqnarray}
The coefficients in these formulas do not decrease monotonically with
increasing their numbers, and the expansion in powers of $w$ is not
alternating having coefficients with irregular signs. Therefore, the RG
series for $\beta$-functions would not demonstrate a good summability
near the chiral fixed point and, subject to Pad\'e-approximant-based
analytical continuation and subsequent Borel integration,
they are hardly believed to yield precise numerical results.
Moreover, the Pad\'e-Borel approximant for $\beta_u$,
taken at the chiral fixed point, as a function of the Borel variable
$t$ has a pole at $t = 61.8$ which is not dangerous practically but
reflects the series poor summability. The difference between numerical
results obtained within RG and MC approaches may be caused by an
unfavorable structure of the RG expansions. On the other hand,
for all $N$ studied the chiral fixed point coordinate $u^*$ given by the
resummed three-loop series remains positive preventing the RG expansions
from losing Borel summability in the domain of interest. Hence,
fortunately, we do not face here this serious problem that may spoil a
perturbative RG analysis, as it occurs, say, when systems with quenched
disorder are investigated \cite{QD}. This keeps calculations of the
higher-order contributions to the RG functions of the model
Eq.(\ref{hamiltonien}) meaningful and desirable.

Can an account for four-loop or higher-order terms in the RG expansions
significantly improve the situation? In principle, yes. The point is that
the exact coordinates of the chiral fixed point may differ substantially
from those given by the three-loop approximation and lie in the domain of
the RG flow diagram where the perturbative expansions of $\beta$-functions
can be properly resummed. Higher-order terms added to the three-loop
series may shift calculated fixed point coordinates toward their exact
values thus making counterparts of the series (\ref{rgexp}-\ref{rgexp2})
better summable. To clear up whether such a situation really takes place,
higher-order RG calculations have to be performed. Until this is done,
an alternative scenario, i. e., the case when higher-order terms do not
improve the summability, cannot be excluded.

There is up to now only one other theoretical apparoach that allows
quantitative calculations in $D=3$: the LPA method that consists in
a truncation of the Wilson RG equations.
Note that even if the LPA is missing the field renormalization and thus
the anomalous dimension $\eta$, it is non
perturbative since it is not based on a weak coupling expansion.
However, although in our case the results obtained within this method
are closer to the MC data than those obtained at the three-loop
RG approximation, they show an unexpected
dependence of $\nu$ with $N$ at small $N$. Moreover, used around $d= 2$,
this approach contradicts the perturbative results obtained from
the Non Linear Sigma model that are, in this dimension, well confirmed by
simulations\cite{Southern}.
They are anyway not enough
accurate to draw a conclusion in $d=3$.
Since the LPA is known to be the first
order of a systematic derivative expansion of the effective action,
it is desirable that the next order be computed.

Let us now remark that even if the three-dimensional physics was well
reproduced by our calculations, it would remain that a coherent picture
of the critical thermodynamics of the frustrated systems
would require to understand the discrepancy between the NL$\sigma$
model approach and the Landau-Wilson one. A striking difference between
both approaches is that near two dimensions the low temperature expansion
of the NL$\sigma$ model predicts that a new ``current'' term of the
form $(\phi^* \nabla\phi)^2$ is relevant\cite{Delamotte}. This term
appears to be fundamental since for $N=3$ it allows to find a fixed
point with an $O(4)$ symmetry. Being highly non renormalizable near four
dimensions, it is irrelevant and forgotten. There is thus another scenario
than the numerical unreliability of the three-loop RG approximation,
namely, that the
Landau-Wilson Hamiltonian Eq.(\ref{hamiltonien}) itself is incomplete
in three dimensions (remember that the presence of topological defects
cannot be invoked here since there are no such defects for $N=6$).
As it was suggested for the Abelian Higgs transition, this could be
interpreted as the necessity to have recourse to the NL$\sigma$ model
description and to abandon that of the Landau-Wilson model. Note, however,
that it is very doubtful that the analysis made around two dimensions
can be extended straightforwardly for any $N$ up to $d=3$ since
i) for $N=3$-component spins the $O(4)$ fixed point found in
$D=2+\epsilon$ has been shown to disappear in a non trivial dimension
strictly smaller than three in a closely related model -- the principal
chiral model\cite{Tissier} -- and since
ii) an $O(4)$ behavior has neither been seen experimentally nor
numerically for $N=3$ and $d=3$. Thus, the perturbative analysis of the
NL$\sigma$ model fails also above two dimensions. However, it remains
that a coherent picture of the behavior of frustrated systems for all
$N$ and $d$ should include the results of the NL$\sigma$ model and
therefore explain why and when the current term starts to be relevant
as a function of $N$ and $d$. If this happens to be around $d=3$ for
$N\sim O(1)$, it could perturb the RG $g$-expansion results presented
here and explain why otherwise powerful methods do not work properly
in our case. In any case, we believe that our results for $N=6$
constitute a clear  challenge for the theoretical approaches which is
perhaps not out of reach from higher order calculations and/or improvement
of the LPA method.

\section {Acknowledgments}
This work was supported in part by the Alexander von Humboldt Foundation
(D. L.), the International Science Foundation (A. I. S., Grant p99-943)
and the Ministry of Education of Russian Federation (A. I. S.,
Grant 97-14.2-16). B. Delamotte and D. Loison are grateful
to G. Zumbach for discussions.

\begin{table}[t]
\begin{center}
\begin{tabular}{c|c|c|c|c|c}
\hspace{-10pt}
\begin{tabular}{c}$N$ \end{tabular}
\hspace{-10pt}
&
\begin{tabular}{c}$\alpha$\end{tabular}
\hspace{-10pt}
&
\begin{tabular}{c}$\beta$\end{tabular}
\hspace{-10pt}
&
\begin{tabular}{c}$\gamma$\end{tabular}
\hspace{-10pt}
&

\begin{tabular}{c}$\nu$\end{tabular}

\hspace{-10pt}
&
\begin{tabular}{c}$\eta$\end{tabular}
\hspace{-10pt}
\\
\hline
5  &  0.305 & 0.300 & 1.095 & 0.565 & 0.0632
\hspace{-10pt}
\\
\hline
6  &  0.275 & 0.302 & 1.121 & 0.575 & 0.0507
\hspace{-10pt}
\\
\hline
7  &  0.303 & 0.295 & 1.108 & 0.566 & 0.0421
\hspace{-10pt}
\\
\hline
8  &  0.152 & 0.319 & 1.211 & 0.616 & 0.0355
\hspace{-10pt}
\\
\hline
9  & -0.055 & 0.354 & 1.348 & 0.685 & 0.0325
\hspace{-10pt}
\\
\hline
10  & -0.157 & 0.370 & 1.417 & 0.719 & 0.0305
\hspace{-10pt}
\\
\hline
12  & -0.292 & 0.393 & 1.506 & 0.764 & 0.0273
\hspace{-10pt}
\\
\hline
16  & -0.451 & 0.418 & 1.616 & 0.817 & 0.0226
\hspace{-10pt}
\\
\hline
20  & -0.553 & 0.434 & 1.685 & 0.851 & 0.0192
\hspace{-10pt}
\\
\hline
24  & -0.623 & 0.444 & 1.734 & 0.874 & 0.0167
\hspace{-10pt}
\\
\hline
100 & -0.909 & 0.488 & 1.935 & 0.970 & 0.0046
\hspace{-10pt}
\end{tabular}
\end{center}
\caption{\protect\label{table1}
Critical exponents calculated by RG
}
\end{table}

\vspace{2cm}

\begin{table}[t]
\begin{center}
\begin{tabular}{c|c|c|c|c|c}
\hspace{-10pt}
\begin{tabular}{c}$N$ \end{tabular}
\hspace{-10pt}
&
\begin{tabular}{c}$\alpha$\end{tabular}
\hspace{-10pt}
&
\begin{tabular}{c}$\beta$\end{tabular}
\hspace{-10pt}
&
\begin{tabular}{c}$\gamma$\end{tabular}
\hspace{-10pt}
&
\begin{tabular}{c}$\nu$\end{tabular}
\hspace{-10pt}
&
\begin{tabular}{c}$\eta$\end{tabular}
\hspace{-10pt}
\\
\hline
6&-0.100(33)$^1$&0.359(14)&1.383(36)&0.700(11)&0.025(20)$^2$
\hspace{-10pt}
\end{tabular}
\end{center}
\caption{\protect\label{table2}
Critical exponents obtained by MC.
$^1$calculated with $\alpha=2-d \nu$.
$^2$calculated with $\eta=2-\gamma/\nu$.
}
\end{table}

\end{document}